# The dynamics of magnetic vortex states in a single permalloy nanoparticle.

Dmitry Ruzmetov[†], Venkat Chandrasekhar

*Department of Physics and Astronomy, Northwestern University, Evanston, IL 60208, USA*

(Last update May 1, 2007)

**Abstract**

We demonstrate a novel method allowing the study of the magnetic state dynamics of a single nanoparticle by means of electron transport measurements. Elliptical 550 nm x 240 nm permalloy nanoparticles are wired with non-magnetic leads for magnetotransport measurements in the presence of a radio-frequency (RF) field. Their resistance exhibits sharp jumps due to the anisotropic magnetoresistance even at room temperature. An RF field induces DC voltage across the nanoparticle which can be partially depleted at a certain RF frequency when a magnetic vortex core resonance is present. An application of an additional DC magnetic field eliminates the vortex and reinstates the unperturbed DC voltage level. The vortex core resonance frequencies are found and the smallest resonance widths are estimated to be less than 6 MHz.

[†] Corresponding author: DEAS, Harvard University, 9 Oxford St., Cambridge, MA 02138, USA, tel. 617-496-5471, fax. 617-495-0346, e-mail: dmitry_ruzmetov@harvard.edu

  

**Introduction**

Nanosize magnets are essential constituents of spintronic devices and promise to be used in future magnetic storage elements. It has been shown that nanomagnets can be in a vortex state with in-plane magnetic structure and a central out-of-plane core [1]. There has been considerable amount of research done on vortex domain structure including theoretical modeling [2], magnetic force microscopy measurements of the cores [3], and vortex dynamics studies [4, 5, 6]. Experimental studies of the magnetic vortex state of a single nanoparticle are a challenge due to the small value of the signal; consequently, most of the research on magnetic vortices has been done with microscopy techniques or on ensembles of particles. We show in this paper that electrical measurements on a single nanoparticle can be used to obtain information about the dynamics of the vortex states of the nanoparticle.

**Experimental procedure**

Three-level electron beam lithography (EBL) using a JEOL JSM-840 scanning electron microscope (SEM) was used to produce the Ni-Fe (permalloy) nanoparticles, Au wires and contact pads on oxidized Si substrates. Ni-Fe nanoparticles and alignment marks were patterned and deposited in the first level of lithography. Fine and low precision gold wires were made in the last two lithography levels. The metals were deposited in an e-gun evaporator on top of the patterned e-beam sensitive resists with subsequent lift-off. Connections from the gold pads on the Si chip to the sample stage were done using a

    

conventional wire-bonder. The sample stage, electrical wires, and the preamp were enclosed in metallic containers for screening purposes. The sample resistance was balanced by a bridge resistance and the small deviations from the constant bridge resistance were amplified and measured. The primary amplification electronics, including the bridge and a 500 gain preamplifier powered by a DC battery, were placed inside a μ-metal shielded box. The 4-wire resistance measurements were done using a lock-in amplifier. The DC voltage on the nanoellipse was amplified by the same 500 gain preamp and measured with an HP multimeter. The sample stage was placed in a brass container between the poles of a DC electromagnet. The radio frequency (RF) signal was supplied by an HP function generator in the range f=0 to 20 GHz.

**Results and Discussion**

Fig. 1a shows an SEM image of 2 devices used in the measurements. Each device consists of an elliptical Ni-Fe 550nm x 240nm particle (Fig 1b) and gold wires defined by EBL. The thin, 100 nm wide wires coming to the particle allow 4-wire resistance measurements. The thick wires above each nanoparticle (Fig 1a) are designed to carry a current from an RF generator. The current creates an RF field at the location of the particle. The measurements were performed on one particle at a time. Both particles showed similar behavior, but the results presented below were obtained on the left particle in Fig 1a whose detailed image is given in Fig 1b.

Fig. 2 shows typical 4-wire resistance measurements on a 550nm x 240nm particle. At small external magnetic fields |B| < 50mT, we observe two distinct resistance states at 13.59 Ohm and 13.56 Ohm (Fig. 2a). These can be explained by the anisotropic magnetoresistance effect (AMR), i.e., the change of electrical resistance due to the variation of the angle between an electric current and magnetization of the sample. As is seen in Fig. 1b, the current in the ellipse flows approximately along the major ellipse axis. Due to the demagnetization field the magnetic moment (**m**) of the ellipse is in the sample plane and is roughly along the major ellipse axis. The two values of the resistance at low fields indicate that there are two equilibrium magnetic states with different angles of the moment with respect to the major axis and, consequently, current. When the field along the major ellipse axis reaches 50mT (Fig. 2a), the magnetic moment switches to align along the external field causing the resistance to jump up. Since the moment is fully aligned there is no more rotation upon further increase of the field resulting in constant resistance above 50mT.

Application of the field perpendicular to the sample surface rotates the moment out of plane, continuously increasing the angle between **m** and the current and causing the resistance to decrease gradually (Fig 2b). It has been shown [7] that such R vs. B data taken at $T = 1.5$ K can be used to reconstruct the magnetic hysteresis curve in a nanoparticle. Our measurements demonstrate the ability to observe sharp resistance jumps due to the magnetic moment switches in a nanoparticle even at room temperature. In this paper we utilize the ability of electrical measurements to extract the information

   

about the magnetic state of a nanoparticle and study the magnetic dynamics of the particle in an external RF magnetic field.

The 4-wire resistance measurements were performed at various frequencies from 17Hz to 150Hz measured by a lock-in amplifier. The lock-in frequency was varied for some measurements to rule out a possibility of instrumental errors and to improve the signal to noise ratio. We also directly monitored the dc voltage across the ellipse. The dc voltage difference ($V_{dc}$) between the bridge and the sample resistances, which is equivalent to the actual voltage resistance across the ellipse, was amplified by a 500 gain preamp inside a shielded μ-metal box and measured by an HP multimeter. The oscillating lock-in component of the voltage across the ellipse was eliminated from such a measurement. Conversely, $V_{dc}$ was filtered out by a lock-in amp in the R vs. B measurements shown in Fig. 2. In the absence of the RF field, the DC voltage was on the order of 2 μV which was the offset for the direct (without lock-in) zero frequency measurements. $V_{dc}$ showed no dependence on the DC magnetic field in the absence of the RF field.

The application of an RF magnetic field in the location of the particle did not produce noticeable changes in their R vs. B characteristics. However the RF field caused a substantial increase of $V_{dc}$ which also showed a rich spectrum depending on the frequency of the RF excitation. A part of the $V_{dc}$ spectrum is shown in Fig 3.

In the measurement, the frequency of the applied current in the top wide wire in Fig 1a was spanned from 60 to 240 MHz while the DC voltage on the ellipse was measured. It is worthwhile to emphasize that there is no hard wire electrical connection between the RF generator circuit and the ellipse circuit (Fig. 4), so that the observed DC voltage is the result of the electromagnetic wave coupling of the RF signal and ellipse circuit (i.e. the circuit containing the ellipse and 4-wire connection leads).

To understand where the coupling of the RF and ellipse circuits takes place we performed a number of tests. The measured sample had a set of ellipses located over a 120x100 μm$^2$ area and each ellipse had a wiring similar to the one shown in Fig 1. We applied the RF current to the RF wire right next to the ellipse and to the RF wire on another ellipse (the device on the right in Fig 1a) on the chip located approximately 30 μm away from the ellipse on which $V_{dc}$ was measured. Turning on the RF signal caused $V_{dc}$ to be induced in both cases but the former case produced significantly higher value of $V_{dc}$. Since the only difference in the two $V_{dc}$ measurements is the distance of the on-chip RF wires from the ellipse we conclude that the coupling of the circuits takes place via a magnetic field generated by an RF wire in the location of the ellipse and that most of the RF field is created by the RF wire right next to the ellipse. As another test, we produced an RF field by using a macroscopic coil (as opposed to the on-chip wire defined by EBL) hand-wound around the sample holder. This circuit also generated an RF field at the ellipse perpendicular to the sample plane but the circuit's geometry was radically different from the on-chip RF lead circuit. Nevertheless we observed an induced $V_{dc}$ of smaller magnitude confirming the conclusion that the DC voltage across the ellipse arises due to the RF field at the location of the nanoparticle.

Since the induced $V_{dc}$ depends on the magnitude of the RF field, the dependence of $V_{dc}$ on the frequency shown in Fig. 3 can be explained by the resonant transmission of the RF circuit, i.e. the circuit consisting of the coax cables from the RF generator to the sample, RF wires lithographically defined on the substrate, etc. The geometry and material of the constituents of the RF circuit will result in the current transmitted to the RF wire on the chip to be dependent on the frequency. The current generates an electromagnetic field at the particle and consequently $V_{dc}$. Accordingly, while the general features of the spectrum in Fig. 3 were preserved (such as the resonances around 150MHz) the precise shape of the spectrum varied depending on the cables used to connect the RF generator to the experimental setup.

While we cannot identify precisely the origin of the induced DC voltage, its appearance is not surprising and can be understood as follows. The wavelength of the RF electromagnetic field well exceeds the size of the nanoparticle in the whole range of the frequencies studied up to 20GHz. Therefore at each moment there is a spatially monotonic gradient of electric field across the ellipse (i.e. there are no field oscillations across the ellipse which would average to zero). Imperfect and asymmetric gold contacts over somewhat oxidized permalloy ellipse create weak tunnel junctions with different barrier profiles at either side of the ellipse. That results in the rectifying effect which converts the RF potential across the ellipse into a DC voltage.

The novel features show up in the analysis of the dependence of Vdc on the external uniform magnetic field **B** along the major ellipse axis produced by a DC electromagnet at the sample location in the addition to the RF field. It was observed that at some frequencies of the RF current ($I_{RF}$ in Fig 4) there was a clear hysteretic dependence of $V_{dc}$ on B. An example of such a dependence is shown in Fig 5a for f = 157 MHz. At the same time there was no appreciable field dependence of $V_{dc}$ at other frequencies (for example at f=100MHz, as shown in Fig 5b). In the absence of an RF field $V_{dc}$ was only a few μV with no distinguishable dependence on **B**.

The $V_{dc}$ behavior displayed in Fig 5a at specific frequencies of the RF excitation can be understood as follows. It has been shown that the magnetic state of ferromagnetic nanoparticles may contain vortices. The cores of the magnetic vortices can have low order uniform precession modes which can be excited by an external RF field in the frequency range from roughly 80 to 300 MHz depending on the nanoparticle geometry [4, 5, 8, 9]. At small external DC magnetic fields the nanoellipse shown in Fig. 1 is likely to have a magnetic vortex structure similar to the one described by Buchanan et al. for Ni-Fe ellipses [9]. The vortex configuration in our case, however, may not be necessarily a simple double vortex state as was described in Ref. [9]. Because the ellipse thickness of our particles is 70 nm, 2D models of the magnetic moment distribution may not be adequate to describe the particle. Given the nanometer scale of our ellipses, the presence of a complex multi-vortex state is expected. A certain frequency of RF excitation will produce uniform precession modes of the vortex cores with the consequent resonance absorption of the RF field in the location of the ellipse. As was discussed above, $V_{dc}$ is proportional to the RF magnetic field, so that the core resonance causes a decrease in $V_{dc}$. The application of a DC magnetic field saturates the magnetization and eliminates the vortices, so that the Vdc increases back to the original value. The resistance of the ellipse measured simultaneously with $V_{dc}$ in Fig 5a is shown in Fig 5c for comparison. We see in Fig 5c that the magnetic moment of the particle is saturated for fields above 70 mT. Since such a saturated state does not have a vortex core we do not expect a resonance and a decrease in $V_{dc}$. Indeed, most of the change in $V_{dc}$ happens below 70 mT. We also see in Fig 5c that the magnetic state at small fields is different for positive and negative B depending on the magnetization history. In this case, the core precession mode is likely to be different too, which explains the asymmetry of $V_{dc}$ in Fig 5a (for example at B = +/-12 mT). Since the nanoellipses are the only elements between the poles of the DC electromagnet which are ferromagnetic, all the observed dependence of $V_{dc}$ on B is from the change of the magnetic state of the particle, which made the field sweeps a tool to detect the vortex resonance absorption.

We studied the DC voltage in a wide frequency range of the $I_{RF}$ from 3MHz to 9GHz. $V_{dc}$ decreases to a few μV at frequencies above 1 GHz which is equivalent to the absence of RF field and probably means that the RF signal does not couple to the sample due to poor matching of the RF generator with the load (RF wire near the nanoparticle) above 1 GHz. $V_{dc}$ has a characteristic spectrum for frequencies from 3 to 300 MHz similar to the section of the spectrum shown in Fig 3. We measured $V_{dc}$ vs. B at various points (frequencies) in the spectrum with an emphasis on the major peaks and minima of the spectrum, and concentrating in the spectrum areas where we found the hysteretic $V_{dc}$ dependence on B. There were two parts of the spectrum at f=140 MHz and f=255 MHz where there was a resonant behavior of $V_{dc}$ vs. B. The first and major part around 140 MHz is indicated by the filled area under the spectrum curve in Fig. 3. It is important to note that the shape of the curve in Fig. 3 reflects the transmission spectrum of the RF line, not the resonance spectrum of the vortex core dynamics. This is because the amplitude of the $V_{dc}$ suppression due to the resonant absorption is on the order of $\Delta V_{dc} =$ 10 µV (cf. Fig 5a), much smaller than mV changes in the $V_{dc}$ due to $I_{RF}$ dependence on the frequency.

In the part of spectrum selected in Fig 3 we measured clear increases in $V_{dc}$ with DC magnetic field at frequencies of the RF excitation (+/- 6 MHz, see explanation below): 120, 133, 137, 149, 151, 155, 157 MHz. $V_{dc}$ vs. B curves at the frequencies looked similar to the one displayed in Fig. 5a. If a resonance is present at a certain frequency, higher power of the RF generator output produces higher absolute value of $V_{dc}$ and correspondingly higher absorption amplitude $\Delta V_{dc}$ up to the highest RF output power that we could test. At some frequencies in the filled region in Fig. 3 (120 to 157 MHz) there was no (or only a weak) field dependence of $V_{dc}$ similar to the curve in Fig 5b. While the presence of a RF field generated a DC voltage whose value increased with increasing RF power, the presence of the *resonance* in the dc voltage did not depend on the absolute value of $V_{dc}$ or RF power. For example, at 155 MHz there was a clear resonance both at $V_{dc}$ = 160 µV and 6 mV, whereas at 150 MHz there was no resonance even at $V_{dc}$ = 5 mV. The RF generator used in our experiments has a long-term (day to day) frequency drift of at least 6 MHz, which complicated the vortex resonance line shape ($\Delta V_{dc}$ vs. f) measurement. Nevertheless, since in the region of 150 +/- 5 MHz we measured both the presence and absence of the resonance, we can conclude that the width of a vortex resonance for a single nanoellipse is less than 6 MHz. For an ensemble of Ni-Fe ellipses, 3 µm x 1.5 µm size and 40 nm thick, Buchanan et al. measured vortex resonances at 142 and 188 MHz with resonance widths above 9 MHz [9]. We see that the single ellipse vortex resonance in our experiment is sharper than that. The precision of our frequency measurement, 6 MHz, lets us conclude that there are at least three resonances present in the 120-157 MHz region for our 550 nm x 240 nm, 70 nm thick ellipses. It is worthwhile to note that the major component of the RF field is perpendicular to the sample plane as can be seen in Fig. 1a, whereas the field is commonly in-plane in the reported vortex resonance experiments [8, 9]. The normal field component was dominant also when the RF field in our experiments was generated by a remote on-chip wire and a macroscopic wire wound around the stage as was discussed above. However since the existence of the in-plane field component is not eliminated we cannot distinguish whether the small in-plane component causes the resonance or it is induced by the normal component of the field.

Outside of the section of the spectrum shown in Fig. 3 we found a resonance only near f = 255 MHz. The shape of the $V_{dc}$ vs. B curve again resembled the one in Fig 5a. Due to our limited time we measured field sweep curves only at a finite number of frequencies, so that we cannot exclude a possibility of the existence of other vortex resonances that we missed. The number of resonances discovered (at least 4) and their frequencies somewhat differ from those reported by other groups [8, 9] and may be due to the differences in the particle sizes, magnetic field orientation, and experimental conditions.

**Conclusions**

We have presented a novel method of studying magnetic dynamics in ferromagnetic nanoparticles by means of electrical measurements in the presence of DC and RF magnetic fields. The advantage of our method is the ability to probe the magnetic dynamics in a single nanoparticle, whereas most of the research in the area is done on ensembles of particles. Non-magnetic (gold) hard wire connections to the nanoparticles ensure direct measurements with the non-perturbed magnetic state of the particle. We observed sharp resonances with widths below 6 MHz of the magnetic vortex core precession in 550 nm x 240 nm Ni-Fe particles. The frequency positions of the resonances are roughly in the same ranges as in the studies on ensembles of similar particles. However the precise frequency positions and number of resonances found are yet to be described by new models which should probably take into account the 3D character of our thick (70nm) particles.

**Acknowledgements**

The work was supported by the Office of Research at Northwestern University. The authors are thankful to Prof. Ketterson for providing the RF function synthesizer.

**Figures.**

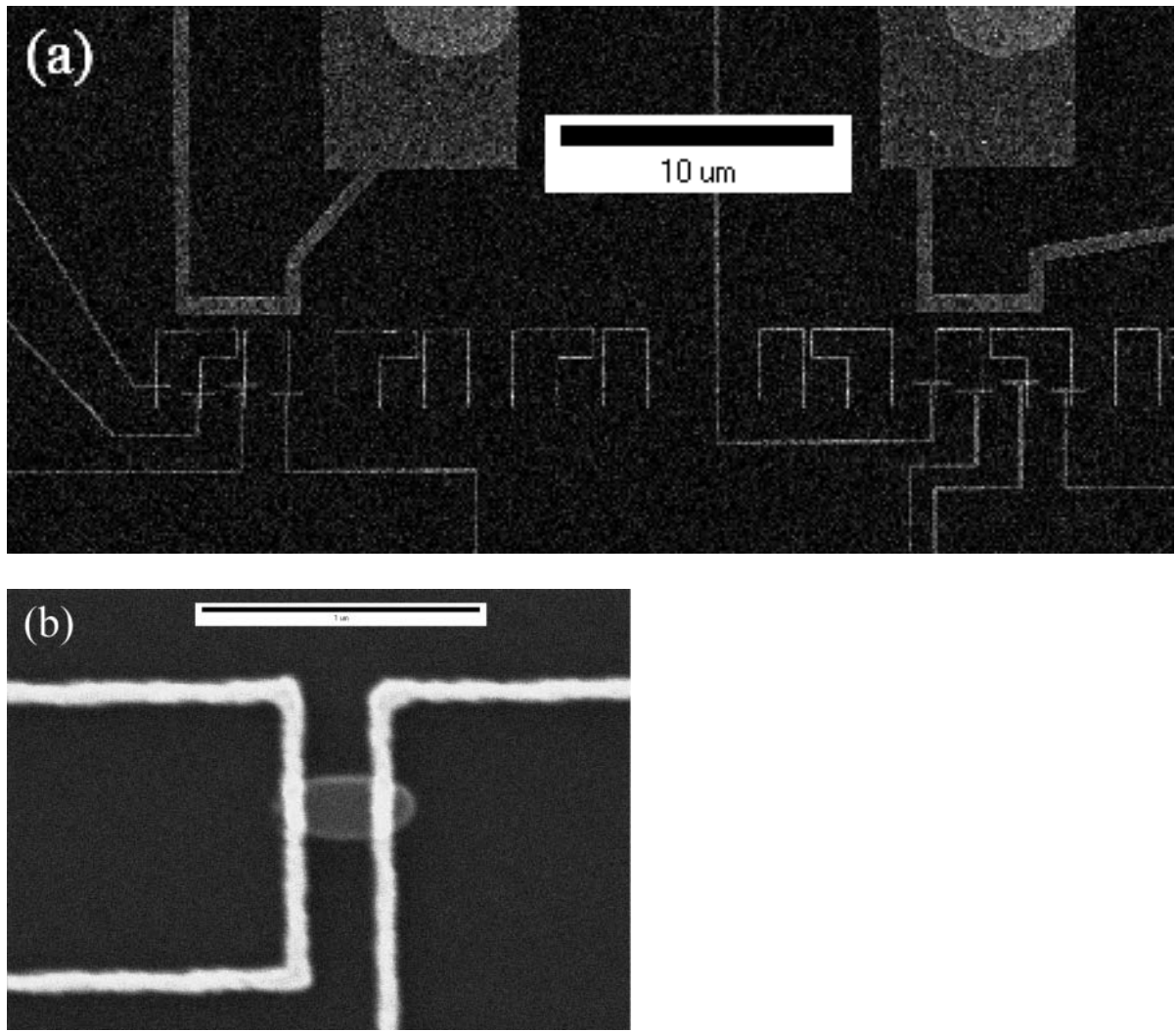

**Figure 1. SEM images of the sample with elliptical Ni-Fe particles and Au wires.**

**(a)** Two identical particles are connected for 4-wire resistance measurements with thin (100 nm) gold leads. Thick wires above the particles are designed to carry RF current. The size bar is 10 μm. **(b)** Expanded view of the left particle in image (a) on which the measurements presented below are done. Dark grey 550 nm x 240 nm particle in the center is made of Ni-Fe, bright wires are Au. The size bar is 1 μm.

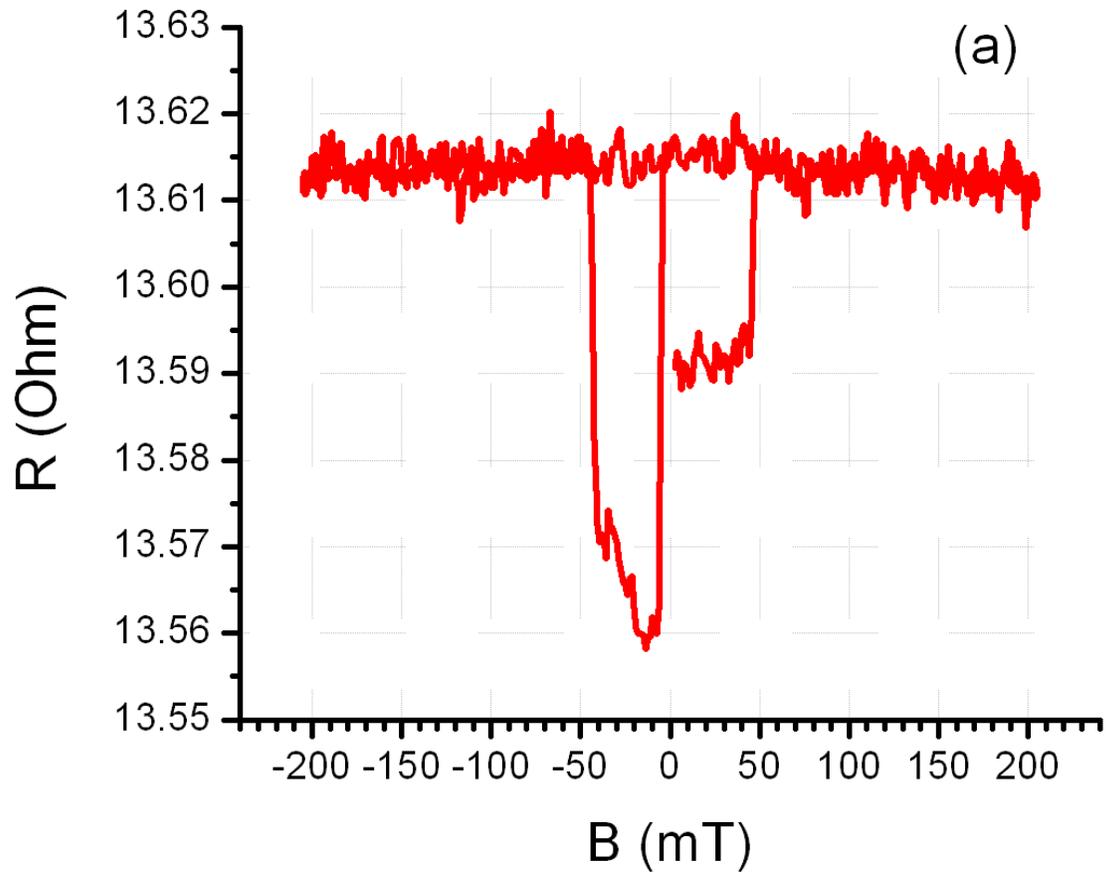

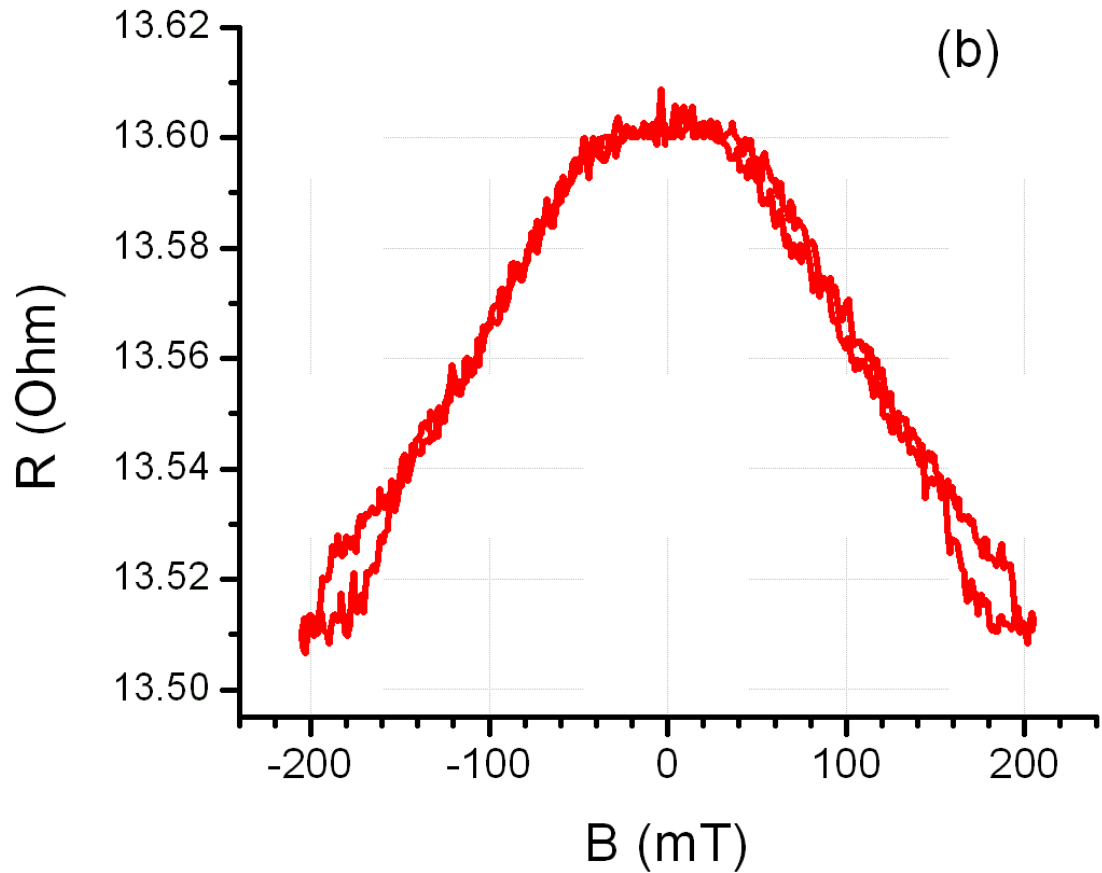

**Figure 2.** 4-wire resistance of an elliptical nanoparticle 550 nm x 240 nm as a function of a DC magnetic field parallel to the major ellipse axis (a) and perpendicular to the sample plane (b).

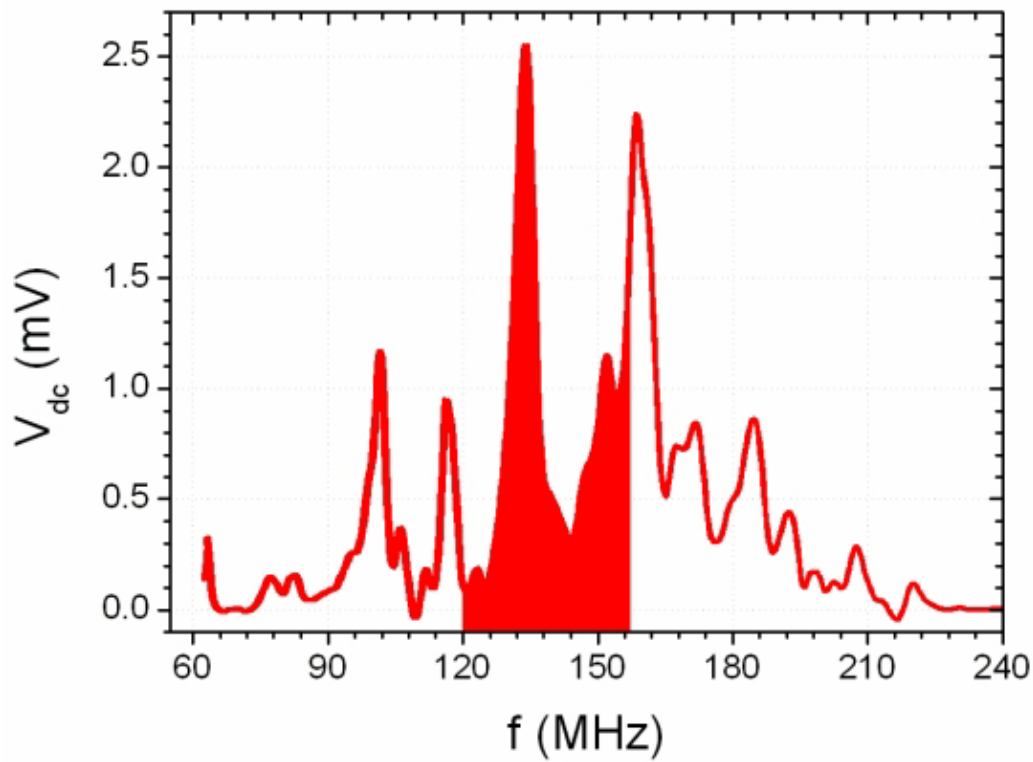

**Figure 3. A part of the spectrum of the DC voltage induced by an RF field on the nanoellipse.** The filled area under the curve indicates the region where the resonance absorption was found at a number of frequencies. The shape of the curve reflects the transmission spectrum of the RF line, not the resonance spectrum of the vortex core dynamics.

   

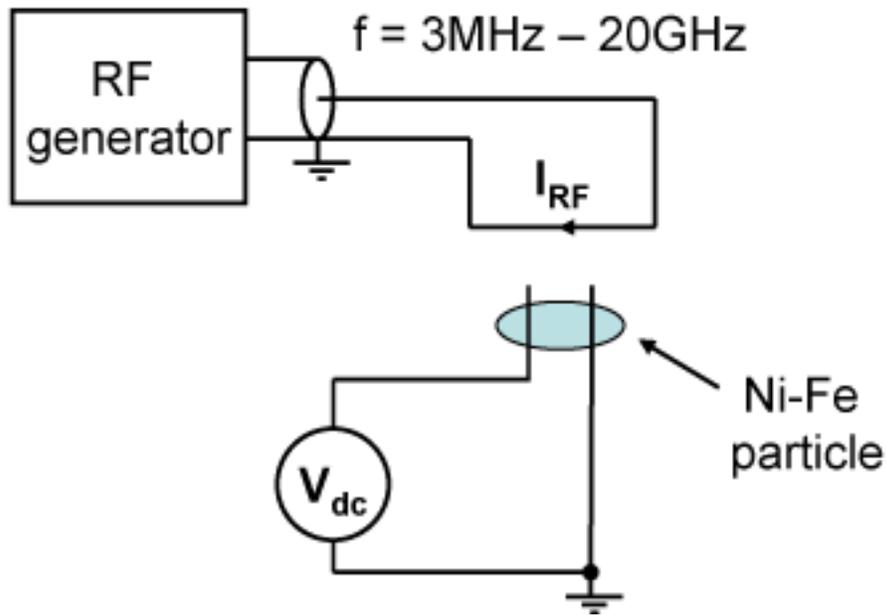

**Figure 4. Simplified electrical diagram of the DC voltage measurement across the nanoellipse.** The RF circuit is electrically isolated from the nanoparticle and the measurement system. The current $I_{RF}$ creates a field at the particle. A preamp, the electronics for 4-wire resistance measurements, and a magnet creating a uniform DC field at the nanoparticle location are not shown.

 

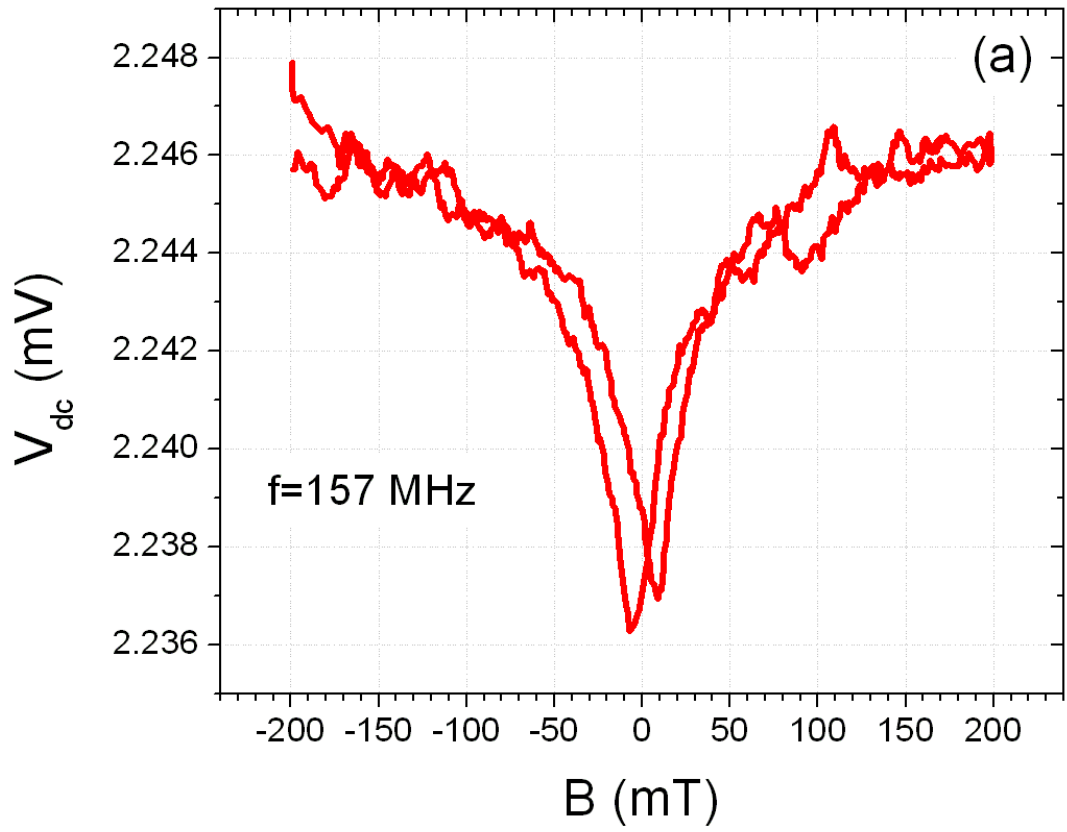

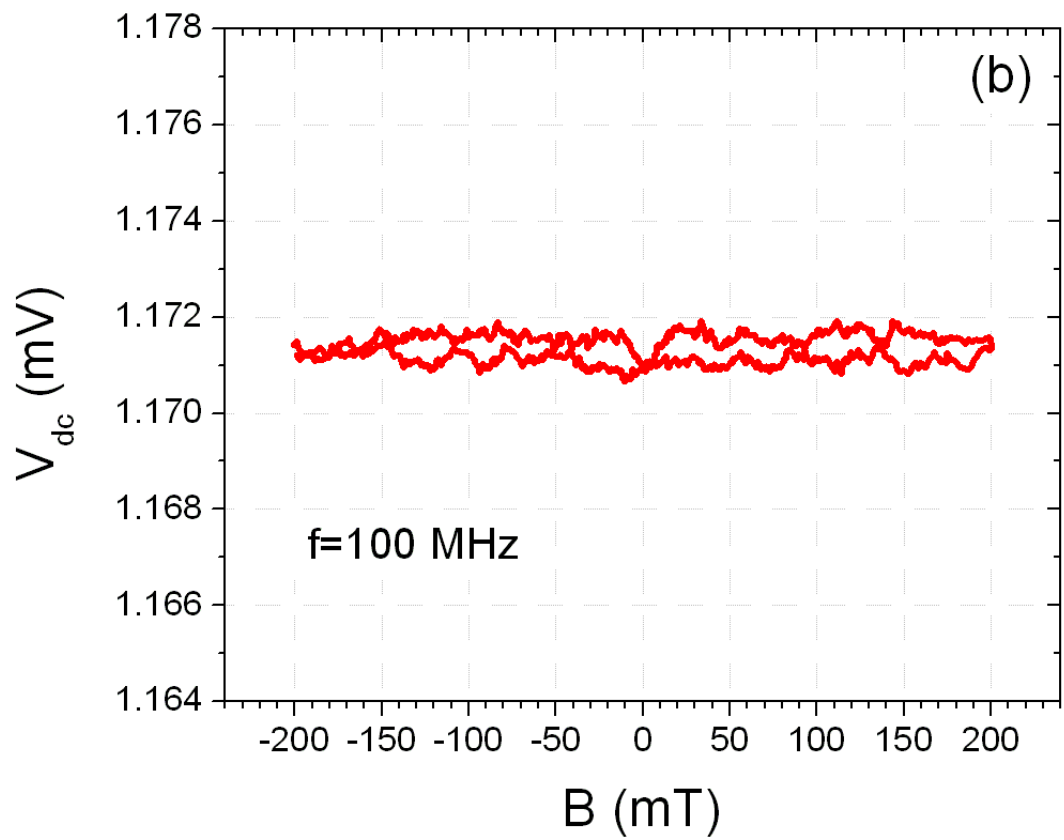

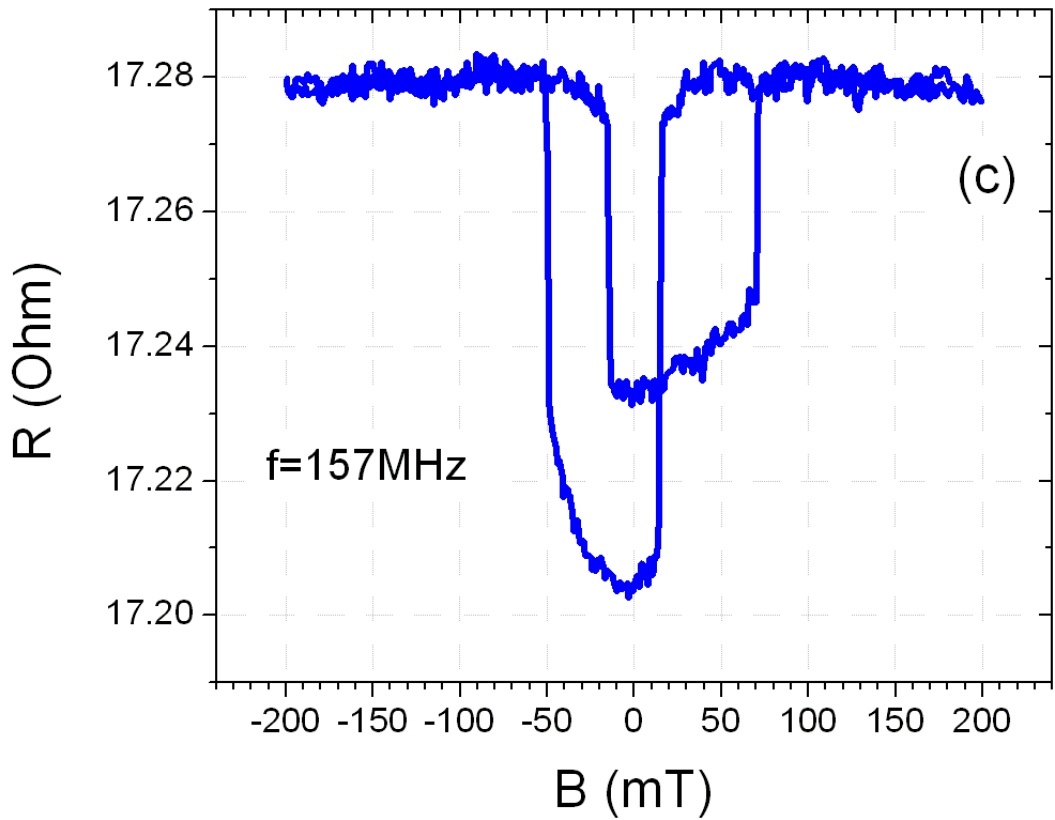

**Figure 5. Magnetic field dependence of the induced DC voltage at different frequencies.** **(a)** DC voltage as a function of a magnetic field generated by a DC electromagnet along the major ellipse axis showing hysteretic behavior (or resonance, see text). The nanoparticle is subjected to an RF field at f = 157 MHz. **(b)** The same measurement as in (a) but at f = 100 MHz. The $V_{dc}$ range is 14 µV equal to the range in (a). There is no dependence of $V_{dc}$ on the field. **(c)** 4-wire resistance measurement of the ellipse using lock-in amplifier in the presence of an RF field at f = 157 MHz.